\documentclass[twocolumn,superscriptaddress,amsmath,aps,prl]{revtex4-2}

\usepackage{graphicx}
\usepackage{dcolumn}
\usepackage{bm}
\usepackage{soul}
\usepackage{xcolor}

\begin{document}

\preprint{APS/123-QED}

\title{Unifying Kappa Distribution Models for Non-Equilibrium Space Plasmas: A Superstatistical Approach Based on Moments}

\author{Abiam Tamburrini}
\email{abiamtamburrini@gmail.com}
\affiliation{Dipartimento di Fisica, Universitá della Calabria, Rende I-87036, Italy}%
 
\author{Sergio Davis}%
\email{sergio.davis@cchen.cl}
\affiliation{%
Research Center in the Intersection of Plasma Physics, Matter and Complexity ($P^2$mc), Comisión Chilena de Energía Nuclear, Casilla 188-D, Santiago, Chile
}%
\author{Pablo S. Moya}%
\email{pablo.moya@uchile.cl}
\affiliation{Physics Department, Faculty of Science, University of Chile, Santiago 8370459, Chile.
}%

\date{\today}

\newcommand{\SD}[1]{\textcolor{blue}{\textbf{#1}}}

\begin{abstract}
From the perspective of non-equilibrium statistical mechanics, modeling the velocity distribution of particles in non-equilibrium, steady-state plasmas presents a significant challenge. Under this context, a family of kappa distributions has been widely used to capture the high-energy tails in space plasmas. These distributions deviate from the canonical Maxwell-Boltzmann statistics and vary significantly in their interpretation of the temperature of an out-of-equilibrium system. In this letter, we establish the validity of any kappa distribution from the standpoint of superstatistics. This study unifies these models by introducing a new kappa distribution based on superstatistical parameters, providing a more general and fundamental framework to connect these distributions and the superstatistical temperature of a system. We demonstrate that the general distribution depends on the thermal characteristics of the modeled temperature distribution population. Furthermore, we present a moment-based velocity distribution that bypasses the traditional temperature debate, relying on the velocity moments. Our findings enhance the understanding of kappa distributions and offer a robust model for non-equilibrium space plasmas.
\end{abstract}

\maketitle

\section{\label{sec:level11} Introduction}

Describing particle velocity distributions in collisionless, out-of-equilibrium plasmas remains a central challenge in space physics, both from theoretical and observational perspectives \cite{modeling_1,modeling_2,modeling_3}. Observations reveal that non-thermal processes—such as wave–particle interactions, turbulence, and plasma instabilities—routinely distort velocity distributions away from thermal equilibrium \cite{bale_16,lopez17}. As a result, the statistical properties of space plasmas frequently deviate from canonical Maxwell–Boltzmann or Jüttner distributions \cite{Livadiotis_2018,yoon_19}. This mismatch has motivated the introduction of alternative empirical models, since equilibrium-based distributions are valid only under strict thermodynamic balance \cite{ourabah_20}.

To account for such non-thermal features, generalized velocity distributions—most notably the family of $\kappa$-distributions—have been widely employed\cite{tsallis,BECK_seA}. These functions extend the Maxwell–Boltzmann form and introduce a parameter $\kappa$ that quantifies the prevalence of suprathermal particles through power-law tails \cite{yoon_20,fichtner_21, leubner2004fundamental}. The $\kappa$-distribution framework has often been linked to Tsallis’ nonextensive statistical mechanics, where Q-exponentials emerge from generalized entropy principles and recover the $\kappa$-functional form under suitable transformations\cite{tsallis, leubner2004fundamental,treumann99, milovanov2000functional}. Despite their success in modeling a variety of space plasmas, the theoretical foundation of $\kappa$-distributions remains debated, particularly regarding the interpretation of their parameters\cite{critique,criticca2,critik4,critic5}.

In practice, two main formulations are commonly adopted. The so-called kappa-A\cite{liva_15} distribution assumes a single Maxwellian temperature, leading to a kappa-dependent thermal velocity \cite{Summerthorne,Leubner82}, while the kappa-B \cite{olbert_68,vasyliunas} distribution instead preserves the Maxwellian thermal velocity but introduces a kappa-dependent temperature\cite{lazar_16}. Both formulations have proven useful in specific heliospheric and magnetospheric contexts\cite{Nicolaou_2018,vinas_2005,lazar2020}, yet they embody distinct extensions of the temperature concept\cite{lazar2022kappa}. As a result, the question of how to consistently generalize temperature out of equilibrium remains unresolved, and the physical meaning of kappa-parameters continues to be actively discussed\cite{H.P._2024,hellberg2009comment,10.1063/1.3213389}.

A promising framework for addressing such non-equilibrium distributions is superstatistics, which describes systems with slow spatial or temporal fluctuations in intensive parameters such as temperature or energy dissipation \cite{BECK_seA}. In this approach, each local domain is assumed to follow an equilibrium distribution, while the global distribution emerges from averaging over these fluctuations. Superstatistics thus provides a natural bridge between local equilibrium descriptions and the complex non-thermal features observed in many physical systems.

This framework has particular relevance for plasma physics. Recent theoretical and numerical studies have shown that superstatistical behavior can emerge naturally in collisionless plasmas evolving toward non-equilibrium steady states \cite{davis_19,Davis_2020}. These results highlight superstatistics as a promising foundation for interpreting non-thermal particle distributions in space environments. However, its role in systematically connecting $\kappa$-like distributions with the physical meaning of intensive parameter fluctuations remains to be fully established. This motivates the present work, where we build on the superstatistical framework to clarify and extend the interpretation of $\kappa$-distributions in plasmas.

In this Letter, we build on the superstatistical framework to clarify the status of $\kappa$-distributions in non-equilibrium plasmas. We demonstrate that the kappa distribution to be used-kappa-A and kappa-B models, as well as the kappa-C parametrization introduced here - depends on the thermal characteristics of the temperature distribution of the population to be modeled. Based on this temperature distribution, we suggest the different contexts in which each kappa model should best represent the system. Furthermore, to avoid deciding on the interpretation or definition of temperature as was done before, we propose a distribution function written in terms of its moments. This extends assumptions' simplicity and fundamental nature to model systems such as non-equilibrium space plasmas\cite{DAVIS_22,davis_23}.

\section{Unified Interpretation of Kappa Distributions via Thermal Velocity Definitions}

Given the microstates of a system, it is possible to express the probability density associated with these microstates in terms of the Hamiltonian of the system. This approach involves considering that the probability distribution depends only on the microstates through the information provided by the Hamiltonian, that is,
\begin{equation*}
    P(\Gamma|S)=\rho(\mathcal{H}(\Gamma)),
\end{equation*}
where $\rho$ is the ensemble function, and $S$ is the set of parameters defining the steady state. It was demonstrated that superstatistics dictates that a distribution of microstates must be given by the following expression,
\begin{equation}
    \rho(E)=\int_0^{\infty} d\beta f(\beta) e^{-\beta E}\,,
    \label{seimpose}
\end{equation}
which corresponds to the Laplace transform of the superstatistical weight function $f(\beta)$ defined by

$$f(\beta):=\frac{P(\beta|S)}{Z(\beta)}\,,$$
where it is noteworthy for its proper formulation, including the partition function $Z(\beta)$. In contrast to other formulations of superstatistics, known as type-A \cite{BECK_seA}, that lead to inconsistencies in the rules of probability for sum and product.

When we assume that the correlation of a particle in a steady state concerning other particles in an ensemble is given by a linear relationship of the kinetic energy, i.e., $k=\gamma_n+\alpha_n K$, after marginalizing over positions, it is possible to demonstrate that the distribution of this particles is given by

$$p_1(k_1)=p_0\left[1+\left(\frac{\alpha_n}{\gamma_n}\right)k_1\right]^{\frac{3n}{2}-\frac{1}{2\alpha_n}-2}.$$

We observe that it precisely exhibits the structure of kappa distributions commonly used, both the kappa-A and kappa-B distribution, for some specific parameterization of the elements $\alpha, \gamma$  with respect to the spectral index $\kappa$. Moving forward, we will demonstrate how both parameterizations are associated with the interpretation of the superstatistical temperature distribution of the system. This interpretation concerns the ensemble temperature in a superstatistical distribution of the Gamma type. Here, the chi-squared distribution and others emerge as particular cases of this.

Assuming the aforementioned, it is possible to express Eq. \ref{seimpose} in the following manner,
\begin{equation}
    p_1(k_1)=\left(\frac{m}{2\pi}\right)^{3/2}\int_0^{\infty} d\beta P(\beta|S)e^{-\beta k_1}\beta^{3/2}.
    \label{pk1}
\end{equation}

Now, considering the distribution of $\beta$ given the parameters of the system, it is crucial to emphasize that in this parameterization of the $\beta$ distribution, the mean is precisely given by the mean fundamental inverse temperature~\cite{DAVIS_22} 
of the system $\beta_S$, as shown by Eq. \eqref{gamma1}, 

\begin{equation}
    P(\beta|u,\beta_S)=\frac{1}{u\beta_S}\frac{1}{\Gamma\left(1/u\right)}e^{-\frac{\beta}{u\beta_S}}\left(\frac{\beta}{u\beta_S}\right)^{1/u-1}.
    \label{gamma1}
    \end{equation}

Replacing Eq. \eqref{gamma1} into Eq. \eqref{pk1} and solving the integral to construct the gamma function yields,


\begin{align}
p_1(k_1)=\left(\frac{m}{2\pi}\right)^{3/2}\left(\frac{1}{u\beta_S}\right)^{-3/2}&\frac{\Gamma\left(1/u+3/2\right)}{\Gamma\left(1/u\right)}\nonumber\\&\left[1+ u\beta_S k_1\right]^{-\left(\frac{3}{2}+\frac{1}{u}\right)}.
\label{p1k1Mod}
\end{align}

This equation represents the velocity distribution in terms of the elements that define the temperature distribution, namely $ u$ and $\beta_S$. These parameters represent the relative variance and the mean of the superstatistical temperature, respectively.

Now, if we specifically choose a parameterization concerning the index, $\kappa$ and $v_{\text{th}}=\sqrt{2/m\beta_{\text{th}}}$ in the following manner,
\begin{subequations}
\begin{align}
\kappa & = \frac{1}{u} + \frac{1}{2}, \\
\label{eq:betaSdef1}
\beta_{\text{th}} & = \frac{\kappa-3/2}{\kappa-1/2}\beta_S,
\end{align}
\end{subequations}
it is possible to rewrite Eq. \eqref{p1k1Mod} as
\begin{align}
    P(v|\kappa, v_{\text{th}_A})=\frac{n}{\pi^{3/2} v_{\text{th}_A}^3}&\left(\kappa-3/2\right)^{-3/2}\frac{\Gamma\left(\kappa+1\right)}{\Gamma\left(\kappa-1/2\right)}\nonumber\\&\left[1+ \frac{1}{\kappa-3/2}\frac{v^2}{v_{\text{th}_A}^2}\right]^{-(\kappa+1)}.
    \end{align}
This precisely corresponds to the modified kappa called kappa-A. Rewriting \eqref{eq:betaSdef1} as,
\begin{equation}
\beta_{\text{th}_A} = \frac{\kappa-\frac{3}{2}}{\kappa-\frac{1}{2}}\beta_S = (1-u)\beta_S,
\end{equation}
we can read $\beta_{\text{th}_A}$ as the mode of the probability density $P(\beta|u, \beta_S)$ in Eq. \eqref{gamma1}.

Also considering the particular parameterization for $\beta_S$ and $\kappa$ as follows,

\begin{equation}
\beta_{\text{th}_B} = \frac{\kappa-\frac{1}{2}}{\kappa}\beta_S = (1+\frac{u}{2})\beta_S,
\label{btholbert}
\end{equation}
and if keep the definition for a thermal velocity $v_{\text{th}_B}=\sqrt{2/(m\beta_{\text{th}_B})}$, we obtain

\begin{align}
P(v|\kappa, v_{\text{th}_B})=\frac{n}{\pi^{3/2} v_{\text{th}_B}^3}&\kappa^{-3/2}\frac{\Gamma\left(\kappa+1\right)}{\Gamma\left(\kappa-1/2\right)}\nonumber\\&\left[1+ \frac{1}{\kappa}\frac{v^2}{v_{\text{th}_B}^2}\right]^{-(\kappa+1)}.
\end{align}
That corresponds to the kappa-B parameterization, but from Eq. \eqref{btholbert}, we note that it is not possible to relate $\beta_{\text{th}_B}$ with any significant statistical quantity in the distribution for the superstatistical temperature, which could give some physical interpretation of this parameterization.

Similarly, we may take
\begin{equation}
\beta_{\text{th}_C} = \beta_S,
\end{equation}
where now $\beta_{\text{th}_C}$ it is linked with the mean of the superstatistical distribution of temperatures and keeps the same definition of $v_{\text{th}_C} = \sqrt{2/(m\beta_{\text{th}_C})}$, we obtain
\begin{equation}
P(v|\kappa, v_{\text{th}_C}) \propto \left[1 + \frac{1}{\kappa-1/2}\frac{v^2}{v_{\text{th}_C}^2}\right]^{-(\kappa+1)}\,,
\end{equation}
which could be a new parameterization of Kappa-type, which we call kappa-C, distribution with a different physical meaning that will be discussed in the next section.

\begin{figure}
    \centering
    \includegraphics[scale=0.4]{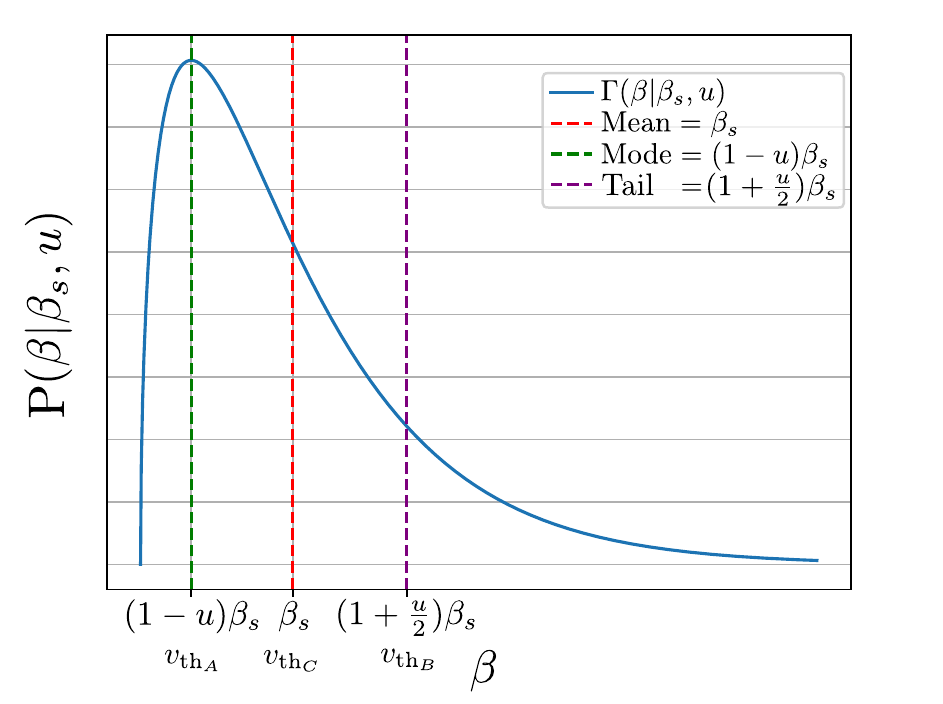}
    \caption{
Superstatistical temperature distribution (blue solid line), indicating each interpretation of temperature associated with thermal velocity for kappa-A parameterization (green dashed line), corresponding to the mode of the distribution, kappa-C (red dashed line), corresponding to the case when the thermal velocity temperature is the mean, and kappa-B (purple dashed line).}
    \label{fig:gamma}
\end{figure}

Each of these interpretations of thermal velocity, which in turn is associated with the concept of temperature achieved in equilibrium by a system, can be observed in Fig.\ref{fig:gamma}. For kappa-A, the temperature value associated with thermal velocity corresponds to the mode of the distribution, i.e., the most probable thermal velocity. Kappa-C's interpretation is linked to the mean of the distribution, while kappa-B, considering fewer cases, is associated with the tail of the distribution. If the relative variance of the temperature distribution is greater, the value associated with kappa-B will shift further towards the outer region of the distribution, representing fewer cases over time.\\  

It is necessary to mention that the parameterization used in the Eq. \eqref{gamma1}, has particularity that the parameter $\beta_S$ coincided with the mean of the distribution. This is not a casualty and response to the fact that this acceptable definition for the temperature of the system is in agreement with the other temperature estimator\cite{rugh,rickayzen}.

On another note, it is important to highlight the definition employed by Lazar \cite{lazar2022kappa} and others to elucidate a potential concept of $\kappa$-dependent temperature. In this context, kinetic temperature is defined as the second moment of a kappa-type distribution. 
$$T^K=\frac{m}{2k_B}\int d{\mathbf{v}} v^2F_K(v)=\frac{m}{2k_B}\frac{\kappa}{\kappa-3/2}v_{\text{th}}^2\,,$$
that if rewrite this in terms of $\beta_S$ we find the relation as follows,

$$\beta^K=\frac{\kappa-3/2}{\kappa}\beta_{\text{th}}.$$
It is possible to analyze this quantity depending on the interpretation given to the temperature associated with thermal velocity. In the case of kappa-A, it would result in a kappa-dependent thermal velocity, so the temperature would not depend on the kappa index. In the case of the parameterization given by kappa-B, it would result in a fundamental kappa-dependent temperature. But in this interpretation, a problem is mentioned by \citet{liva_15}.

Certainly, when calculating the second moment of the general distribution provided in Eq. \eqref{p1k1Mod}, it is feasible to express it in terms of the 'kinetic temperature' as follows:

$$\beta^K=\frac{2}{3} \beta_S(1-u).$$
This definition aligns seamlessly with both the mode of the temperature distribution and the definition of thermal temperature in the Kappa-A parametrization,

$$\beta^K=\frac{2}{3}\beta_{\text{th}_A}.$$
Now, we have demonstrated that it is possible to obtain the parameterization under the assumption that the temperature associated with the thermal velocity of the velocity distribution corresponds to the mode of the temperature distribution. Consequently, there is no need for the justification provided by Livadiotis regarding inheriting the temperature definition as the quantity that makes sense as soon as the kinetic definition of Maxwell and the thermodynamic definition of Clausius coincide for systems out of equilibrium. Therefore, this does not invalidate the expression given by the kappa-B parameterization, even without finding a significant statistical quantity in the temperature distribution for the temperature associated with its thermal velocity definition. However, we note from Olbert's explanation in 1968 for this kappa distribution that he refers to what we interpret today as thermal velocity. Specifically, the most probable velocity, which, in this case, would be associated with the thermal velocity definition given by the mode of the temperature distribution. This coincides with the kappa-A expression.

\section{Moment-Based Formulation of Velocity Distributions}

The concept of temperature arises from thermodynamics when systems are in equilibrium. However, extending this definition or considering other definitions for non-equilibrium systems proves nontrivial and often loses validity in different scenarios. As demonstrated in the previous section, considering the uncertainty in the equilibrium temperature allows us to revert to a distribution with a certain degree of uncertainty. The definition derived from the thermal velocity of a particle distribution gives rise to different ways of modeling these distributions, depending on the interpretation given. At least three forms of particle distribution can be reached, each representing different characteristics within the population. All three should be able to fit well for different cases, implying that all these distributions could fit well for various plasmas with special characteristics. Therefore, since the form of the distribution depends on the chosen definition of temperature in thermal velocity, all options are admissible and valid under the framework of Superstatistics. 

Now, to avoid making decisions regarding the definition of temperature, let us propose going further and having a much more general expression of the particle distribution, where the previous cases are included as special cases. This more general form depends only on the moments of the distribution, not necessarily all of them. Because the particle distribution, as formulated by Superstatistics, depends on variables to be completely defined, we will need, until now, two moments of the distribution to construct it.

The mean and relative variance of $P(k_1|u, \beta_S )$ in Eq. \eqref{p1k1Mod} are
given by
\begin{equation}
    \left\langle v^2\right\rangle \sim \left\langle k\right\rangle_{u,\beta_S} = \frac{3}{2}\frac{1}{\beta_S(1-u)},
    \label{eq:v2}
\end{equation}

\begin{equation}
    \frac{\left\langle (\delta k)^2\right\rangle_{u,\beta_S}}{\left\langle k\right\rangle_{u,\beta_S}^2} = \frac{2+u}{3(1-2u)}.
    \label{eq:relativvariancek}
\end{equation}

From these two equations, it is possible to find the fourth moment of the velocity distribution and, with that, the kurtosis,

\begin{equation}
    \frac{\left\langle k^2\right\rangle_{u,\beta_S}}{\left\langle k\right\rangle_{u,\beta_S}^2}=\mathcal{K} = \frac{5(1-u)}{3(1-2u)},
    \label{eq:kvsU}
\end{equation}
considering the definition of kurtosis in the form,
\begin{equation}
    {\text{Kurt}}\left[v\right]=\mathcal{K}=\frac{ \left\langle v^4\right\rangle_{u,\beta_S}}{ \left\langle v^2\right\rangle_{u,\beta_S}^2}.
    \label{eq:kurt_1}
\end{equation}
For the Maxwellian case, it is possible to compute the moments in terms of the dimension as follows,

$$\langle v^l\rangle=\frac{(l+d-2)!!}{(\beta m)^{l/2}}.$$
Then we compute the kurtosis as in Eq. \eqref{eq:kurt_1}, considering $d=3$ as the dimension of the velocity field, we find $\mathcal{K}=\frac{5}{3}$. Which corresponds when taking the limit $u=0$ in Eq. \eqref{eq:kvsU}.



We note that the Eq. \eqref{eq:relativvariancek} must monotonically increase with $u$ and that, given it must remain non-negative, implies that $0 \leq u < 1/2$, which in turn has implications for $\mathcal{K}$. Therefore, from the Eq. \eqref{eq:kvsU}, we can deduce,
\begin{equation}
   u=\frac{3\mathcal{K}-5}{6\mathcal{K}-5}
    \\
    \implies \mathcal{K}\geq \frac{5}{3}.
    \label{eq:uvsk}
\end{equation}
 According to the classification of kurtosis, these distributions correspond to leptokurtic distributions, meaning that, as expected, they exhibit elongated tails compared to the equilibrium case described by a Maxwell-Boltzmann distribution.

To construct a distribution in terms of its moments, we take the average of the kinetic energy from Eq. \eqref{eq:v2} as follows,

\begin{equation}
    \mathcal{E}_m=\left\langle v^2\right\rangle= \frac{3}{m}\frac{1}{\beta_S(1-u)},
    \label{eq:energy}
\end{equation}
and solving with Eq. \eqref{eq:kvsU} for $\beta_S$, we obtain,

\begin{equation}
    \beta_S=\frac{6\mathcal{K}-5}{m\mathcal{K}\mathcal{E}_m}.
\end{equation}
With the set of Eqs. \eqref{eq:energy} and \eqref{eq:kvsU}, we can write the velocity distribution based on Eq. \ref{p1k1Mod} in terms of the mean kinetic energy and kurtosis, as follows:


\begin{align}
P&(v|\mathcal{K}, \mathcal{E}_m)=\left(\frac{1}{2\pi}\right)^{\frac{3}{2}}\left(\frac{\mathcal{K}\mathcal{E}_m}{3\mathcal{K}-5}\right)^{-\frac{3}{2}}\nonumber\\&\frac{\Gamma\left(\frac{6\mathcal{K}-5}{3\mathcal{K}-5}+\frac{3}{2}\right)}{\Gamma\left(\frac{6\mathcal{K}-5}{3\mathcal{K}-5}\right)}\left[1+  \frac{3\mathcal{K}-5}{2\mathcal{K}\mathcal{E}_m}v^2\right]^{-\left(\frac{6\mathcal{K}-5}{3\mathcal{K}-5}+\frac{3}{2}\right)}.
\label{eq:p1k1Moments}
\end{align}
By construction, this equation spawns any of the three kappa parameterizations without the need to decide on the interpretation of the temperature associated with the thermal velocity.

\section{Discussion}

We have effectively concluded that both models for kappa distributions may be valid in distinct physical systems \cite{lazar2022kappa}. The primary distinctions lie in that the Kappa-A model is associated with a process that enhances the central part of the velocity distribution, potentially linked as a mechanism to an augmented effective ``collision'' rate influenced by wave-particle interactions \cite{lazar_16}. Consequently, it is associated with a potential external source. This is related to the interpretation associated with the temperature distribution in the system. In this case, it is determined by the mode of the temperature distribution comprising the system.

The mean of a distribution is generally linked to collective behaviors, serving as a weighted average that considers the contribution of all particles in the system. Therefore, it is more sensitive to global variations in system velocities. Moreover, it may better reflect the collective response of the system to external influences. It considers the contribution of all different particles in the system. This makes it a predominant mean in a system requiring or sensitive to changes induced by an external source, as exemplified by the Kappa-C model described earlier.

On the other hand, the Kappa-B model would better describe a system where the increase in high velocities is more predominant due to a lack of collisions. In this context, it makes sense to obtain an interpretation of the temperature distribution associated with the tail of higher temperatures, as shown in Fig. \ref{fig:gamma}. Which elevates the tails much more than other interpretations, making them an important value in this system. It would be associated with predominant rapid events in these phenomena. Similarly, the mode can provide a more localized view of the distribution, relevant for more specific phenomena without being affected by extreme velocities in other parts. These findings contribute to a deeper understanding of the interpretations associated with kappa distributions based on the fundamental characteristics of the considered physical systems.

A velocity distribution based on its moments has also been presented to avoid deciding how the concept of temperature should behave out of equilibrium, thus bypassing this discussion for using the different kappa models. In principle, this distribution depends on the second and fourth moments to determine the two parameters that define the temperature distribution. 
This reliance on even moments arises because the superstatistical expression is based on a symmetric formulation. Future work includes incorporating asymmetry into the distribution. 
 This approach could encompass more non-thermal characteristics observed in the solar wind, such as asymmetries caused by heat flux instabilities \cite{bea22}. By accessing the odd moments of the distribution, we could analyze how they influence the temperature distribution. We believe this could potentially describe these phenomena through generalized anisotropic superstatistics, generating peculiarities in the temperature distribution as introduced by Sánchez \emph{et al}\cite{sanchez}. This framework not only captures these phenomena through generalized anisotropic superstatistics but also paves the way for further exploration of temperature distributions in complex systems.

\section{Acknowledgments}
\begin{acknowledgments}
We gratefully acknowledge funding from ANID, Chile, through FONDECYT grants No. 1220651 (S.D.) and No. 1240281 (P.S.M). A. Tamburrini is grateful to the Agencia Nacional de Investigación y Desarrollo (ANID, Chile) for the National Doctoral Scholarship No. 21210407, and to the Space It Up project, funded by the Italian Space Agency (ASI) and the Ministry of University and Research (MUR), under Contract No. 2024-5-E.0 – CUP No. I53D24000060005.
\end{acknowledgments}

\bibliography{apssamp}
 \newpage

\end{document}